# Quantum critical state in a magnetic quasicrystal

Kazuhiko Deguchi[1], Shuya Matsukawa[1], Noriaki K. Sato[1],

Taisuke Hattori[2], Kenji Ishida[2], Hiroyuki Takakura[3] & Tsutomu Ishimasa[3]

[1]*Department of Physics, Graduate School of Science, Nagoya University, Nagoya 464-8602, Japan*

[2]*Department of Physics, Graduate School of Science, Kyoto University, Kyoto 606-8502, Japan*

[3]*Division of Applied Physics, Graduate School of Engineering, Hokkaido University, Sapporo 060-8628, Japan*

**Quasicrystals are metallic alloys that possess long-range, aperiodic structures with diffraction symmetries forbidden to conventional crystals. Since the discovery of quasicrystals by Schechtman *et al.* at 1984 (ref. 1), there has been considerable progress in resolving their geometric structure. For example, it is well known that the golden ratio of mathematics and art occurs over and over again in their crystal structure. However, the characteristic properties of the electronic states – whether they are extended as in periodic crystals or localized as in amorphous materials – are still unresolved. Here we report the first observation of quantum ($T = 0$) critical phenomena of the Au-Al-Yb quasicrystal – the magnetic susceptibility and the electronic specific heat coefficient arising from strongly correlated 4$f$ electrons of the Yb atoms diverge as $T \rightarrow 0$. Furthermore, we observe that this quantum critical phenomenon is robust against hydrostatic pressure. By contrast, there is no such divergence in a crystalline approximant, a phase whose composition is close to that of the quasicrystal and whose unit cell has atomic decorations (that is,**

**icosahedral clusters of atoms) that look like the quasicrystal. These results clearly indicate that the quantum criticality is associated with the unique electronic state of the quasicrystal, that is, a spatially confined critical state. Finally we discuss the possibility that there is a general law underlying the conventional crystals and the quasicrystals.**

The quasicrystal that we study here is a gold-aluminum-ytterbium alloy described as $Au_{51}Al_{34}Yb_{15}$ with a six-dimensional lattice parameter $a_{6d} = 0.7448$ nm. In Fig. 1a we present the electron diffraction pattern of our sample demonstrating 5-fold symmetry. This exhibits the characteristic feature of quasicrystals − long-range translational but aperiodic order. Owing to this quasi-periodicity, an unusual electronic state that is neither extended nor localized is expected. However, such an unusual state has not yet been observed. In the present paper, we show the Au-Al-Yb quasicrystal to present a peculiar quantum critical behaviour, which we propose to reflect this unusual state expected for quasicrystals. It becomes apparent in the present system because of strong correlations induced by the $4f$ electrons of Yb.

The Au-Al-Yb quasicrystal was discovered in the course of research on new series of Tsai-type quasicrystals[2]. The Yb valence was found to be intermediate between $Yb^{2+}$ and $Yb^{3+}$ by means of experiments of X-ray absorption near edge structure (XANES), indicating the hybridization of the $4f$ electrons of the Yb atoms with the conduction electrons. Figure 1b shows the arrangement of Yb atoms in the structure model of the cadmium-ytterbium ($Cd_{5.7}Yb$) quasicrystal[3,4] which is isostructural with the present Au-Al-Yb quasicrystal. For comparison, we illustrate the crystal structure of the approximant $Au_{51}Al_{35}Yb_{14}$ in Fig. 1c. The edge length of the icosahedron ranges from 0.545 to 0.549 nm. In reality, this icosahedron corresponds to the third shell of the Tsai-type cluster (Figs. 1d - h).

In what follows, we attempt to reveal the nature of the characteristic electronic state of quasi-periodic systems by probing the $4f$ electrons and comparing physical properties between the quasi-periodic quasicrystal and the periodic approximant. Before presenting the low temperature data, we briefly mention the high temperature results of the electrical resistivity $\rho(T)$ and the magnetic susceptibility $\chi(T)$. As shown in Supplementary Fig. S1a, both the quasicrystal and the approximant show metallic behaviour in a wide temperature range with a large residual resistivity $\rho(0)$, but they show different behaviour at low temperatures (inset of Supplementary Fig. S1a); whereas the approximant exhibits the conventional Fermi-liquid power-law $\Delta\rho \propto T^2$ (where $\Delta\rho = \rho(T) - \rho(0)$), the quasicrystal rather exhibits the $T$ linear dependence, $\Delta\rho \propto T$. To our knowledge, this is the first observation of the so-called non-Fermi liquid behaviour in quasicrystals. For magnetism, both the quasicrystal and the approximant show a Curie–Weiss form above 100 K (Supplementary Fig. S1b), which yields an effective moment of $\mu_{\text{eff}} = 3.91\mu_B$ and $3.96\mu_B$ for the quasicrystal and the approximant, respectively. These values indicate that the Yb-ion valence of both the quasicrystal and the approximant is in between $Yb^{3+}$ and $Yb^{2+}$.

Now we focus on the quantum criticality (that is, the critical behaviour near $T = 0$) of the quasicrystal. The magnetic susceptibility of the quasicrystal at $H = 0$ shows a divergent behaviour as $T \to 0$ (Fig. 2a), and this quantum criticality is characterized by a critical index $n = 0.51$ as defined by $\chi^{-1} \propto T^n$ (Fig. 2a inset). To examine the effect of pressure on this critical behaviour, we measured the magnetic susceptibility under hydrostatic pressure. The results are indicated in the inset of Fig. 2a. It is clearly seen that the divergent behaviour survives with the novel critical exponent unchanged. Here it should be noted that the hydrostatic pressure can change the hybridization but not alter the crystal symmetry. In contrast, the application of a magnetic field suppresses the divergence, resulting in the saturation of $\chi(T)$ at low temperatures. This allows us to define a crossover temperature $T^*$ into the field-induced Fermi liquid state with the enhanced Pauli susceptibility, as indicated on the figure by the arrow at which the $\chi(T)$ curve shows a maximum. In contrast, the magnetic susceptibility of the approximant increases monotonically with decreasing temperature (Fig. 2b), but does not diverge as evidenced from inset of Fig. 2a, $\chi^{-1} \propto T^n + constant$ with the same exponent $n$ as above.

Figure 3a shows the temperature–field phase diagram based on a contour plot of the normalized uniform susceptibility, $\chi(T, H)/\chi(T, 0)$. We note that $T^*$ (open square) seems to approach zero as $T \to 0$. A crossover field $H^*$ defined by $\chi(T, H^*)/\chi(T, 0) = 0.95$ (open circle) falls down to zero as $T \to 0$, meaning that $\chi(H)$ diverges as $H \to 0$ at $T \sim 0$. These results indicate that there is a singular point at $T = H = 0$ without chemical doping and pressurization. In this respect, the Au-Al-Yb quasicrystal is regarded as a quantum critical matter.

The nuclear spin-lattice relaxation rate divided by temperature, $1/T_1T$, deduced from $^{27}$Al nuclear magnetic resonance (NMR) measurements on the quasicrystal is plotted in Fig. 3b. (The recovery curves are shown in Supplementary Fig. S2: they are fitted well using a single component of $1/T_1$, indicating that the Al sites around the Yb ions are not inhomogeneous.) Whereas the aforementioned uniform susceptibility $\chi(T)$ probes magnetic fluctuations at $q = 0$ (where $q$ is the wave vector of an applied magnetic field), $1/T_1T$ observes the $q$-averaged fluctuations. The scaling observed here, $1/T_1T \propto \chi(T)$, together with a negative Weiss temperature suggest that $\chi(q)$ is independent of $q$, meaning that the magnetic fluctuation associated with the quantum criticality possesses a local nature in the real space.

Let us move onto the heat capacity (Fig. 4). For the quasicrystal, the logarithmic divergence at zero field is observed in the temperature dependence of the magnetic specific heat ($C_M$) divided by temperature, $C_M/T \propto -\ln T$ (Fig. 4a and inset of Fig. 4b). By contrast, the approximant shows no divergence (Fig. 4b), although the saturated value is very large, $\sim 0.7$ JK$^{-2}$mol$^{-1}$, compared to conventional crystals and quasicrystals. In magnetic fields, the divergence of the quasicrystal is suppressed (Fig. 4a), but the saturated value is still very large; $C_M/T \sim 0.2$ JK$^{-2}$mol$^{-1}$ at $H = 50$ kOe. The approximant shows the similar field effect. These results are consistent with the suppression of $\chi(T)$ by the magnetic field, supporting the field-induced Fermi liquid state.

Combining the magnetic and thermodynamic results, in the inset of Fig. 2b we plot the ratio $\chi/\gamma$ (where $\gamma = C_M/T$), which is a measure of the magnetic correlation of quasiparticles. We note that the ratio is enhanced at low fields for both the quasicrystal and the approximant, suggesting the presence of magnetic correlations there.

The experimental results presented above are summarised as follows. Both the periodic approximant and the quasi-periodic quasicrystal show similar transport and magnetic properties at high temperatures. The difference becomes evident at low temperatures: whereas the approximant shows the Fermi liquid behaviour, the quasicrystal exhibits non-Fermi liquid at zero field (see also Supplementary Table S1). We interpret this difference, the presence/absence of the divergence in $\chi$ and $C_M/T$, as the presence/absence of the critical state unique to quasicrystals with the quasi-periodicity[5]. This interpretation is supported by the robustness of the quantum criticality against the hydrostatic pressure: for crystalline materials, a perturbation such as the application of external pressure gives rise to deviation from the critical 'point'. As a result, we conclude that the present quantum criticality is associated with the unique electronic state of quasicrystal.

Finally, we discuss the implication of the robustness against hydrostatic pressure. In general, the quasicrystal critical state can be characterized by an extremely degenerate confined wave function[5-7] and singular continuous density of states[5,8-12]. The robustness suggests that the critical state is also robust against pressure; this is naturally understood because the hydrostatic pressure does not change the symmetry and the quasi-periodicity. For the unusual exponents of the quantum criticality, on the other hand, there are some models that may account for the non-Fermi liquid behaviour. A valence criticality theory successfully accounts for our non-Fermi liquid exponents, as seen in Supplementary Table S1 (ref. 13). Although the theory assumes the crystalline materials, it may get at the essence of the critical state (that is, the spatially local nature of the 4$f$ electrons) by assuming the dynamical exponent $z = \infty$. This suggests the possibility that there is a general law underlying the conventional crystals and the quasicrystals, however the theory is unlikely to explain the robustness. Other theories such as the so-called Kondo disorder and Griffiths phase seem to be also excluded: see Supplementary Information for more detailed discussion. To our knowledge, there is no theory to explain the quantum criticality that is robust against hydrostatic pressure but readily destroyed by magnetic field. We hope that the present results stimulate theoretical works.

**Acknowledgments**

The authors thank Y. Tanaka and S. Yamamoto for supports of the experiments. The authors also thank S. Kashimoto, T. Watanuki, S. Watanabe, K. Miyake and Y. Takahashi for valuable discussions. This work was partially supported by a grant-in-aid for Scientific Research from JSPS, KAKENHI (S) (No. 20224015), the "Heavy Electrons" Grant-in-Aid for Scientific Research on Innovative Areas (No. 20102006, No. 21102510, No. 20102008, and No. 23102714) from MEXT of Japan, a Grant-in-Aid for the Global COE Program ``The Next Generation of Physics, Spun from Universality and Emergence'' from MEXT of Japan, and FIRST program from JSPS.

**Author Contributions**

K. D., T. I., K. I. and N. K. S. wrote the paper. K. D., S. M. and N. K. S. carried out low temperature experiments. K. I. and T. H. carried out NMR experiments. T. I. and H. T. carried out sample preparations and structure determinations. All authors discussed the results and commented on the manuscript.

**Additional information**

Supplementary information is available in the online version of the paper. Reprints and permissions information is available online at www.nature.com/reprints. Correspondence and requests for materials should be addressed to K.D.

**Competing financial interests**

The authors declare no competing financial interests.

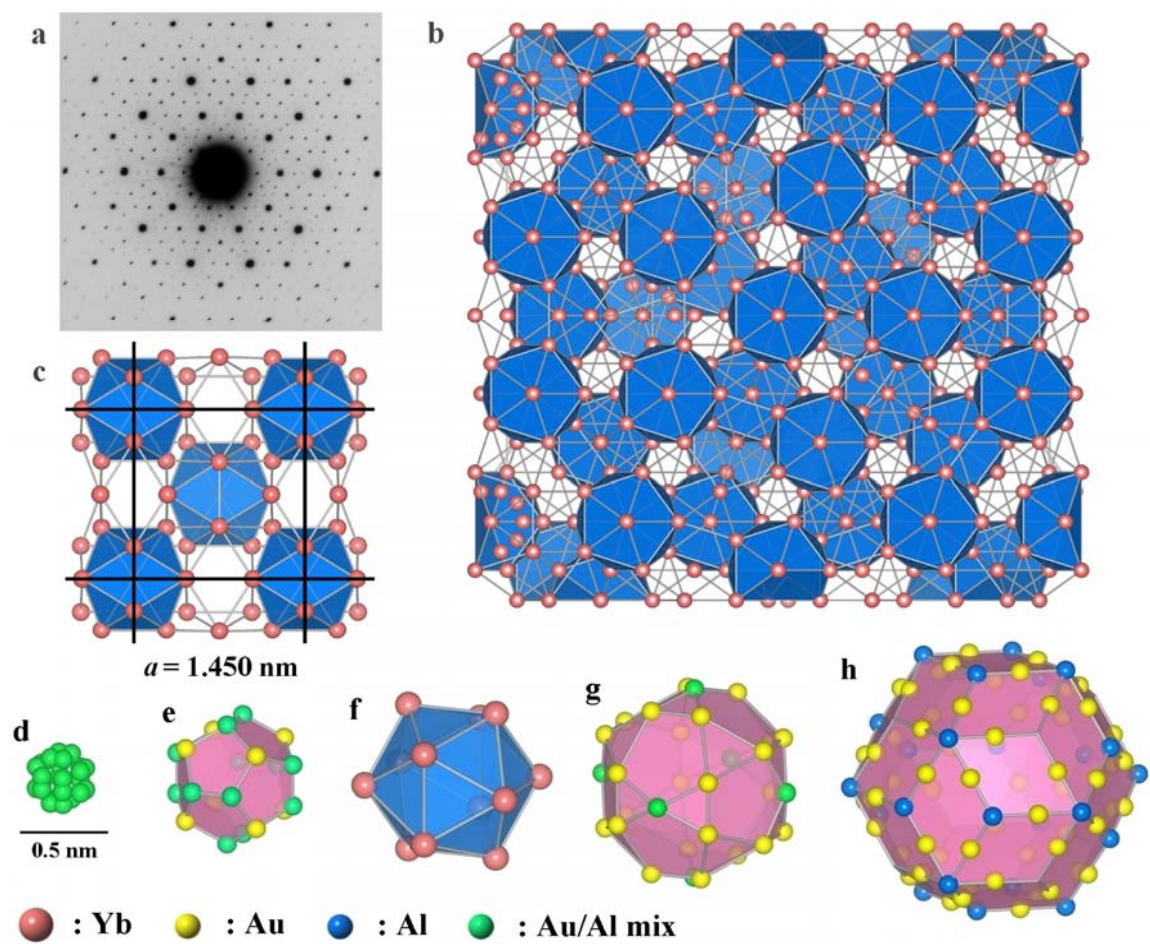

●: Yb  ●: Au  ●: Al  ●: Au/Al mix

**Figure 1 | Structure models of quasicrystal and approximant. a,** Selected-area electron diffraction pattern of the Au-Al-Yb quasicrystal. **b,** Atomic arrangement of Yb atoms in the isostructural $Cd_{5.7}Yb$ quasicrystal. The Yb atoms included in a cube with an edge length of 6 nm are projected onto the plane perpendicular to the 5-fold axis. The icosahedral aggregate is highlighted. **c,** Yb arrangement in the Au-Al-Yb approximant in the projection along the [001] direction. **d-h,** Concentric shell structures of Tsai-type cluster in the Au-Al-Yb approximant. Each vertex of the first cluster presented in **d** is occupied by Au/Al mixed atoms with an occupancy 1/6. Chemical ordering in each shell is based on the result of structure analysis of the approximant (see Ref. 2 for details). Scale bar (0.5 nm) is shown at **d**.

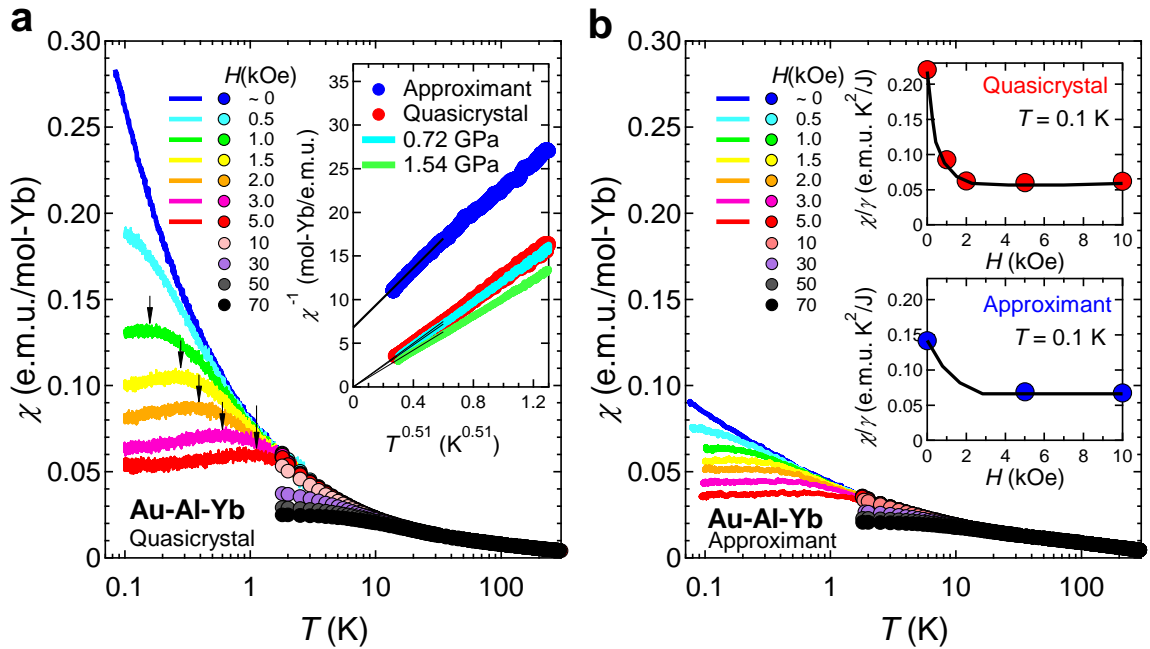

**Figure 2 | Temperature dependence of the magnetic susceptibility of the quasicrystal and the approximant. a**, Ac and dc magnetic susceptibility of the quasicrystal measured in a temperature range of 0.08 < $T$ < 3.0 K (denoted by the line) and 1.8 < $T$ < 300 K (circles), respectively. Magnetic fields are described in the figure. The abscissa is plotted on a logarithmic scale. The arrows indicate a characteristic temperature $T^*(H)$. Inset shows the inverse susceptibility $\chi^{-1}$ versus $T^{0.51}$ of the approximant (blue circles) and the quasicrystal (red circles) at ambient pressure, and of the quasicrystal at pressures of 0.72 GPa and 1.54 GPa. Black lines are linear extrapolation to $T$ = 0 K. **b**, Magnetic susceptibility of the approximant measured in the same condition as in Fig. 2a. Inset shows the field dependence of the ratio $\chi/\gamma$ at $T$ = 0.1 K for the quasicrystal and the approximant, respectively, where $\gamma = C_M/T$.

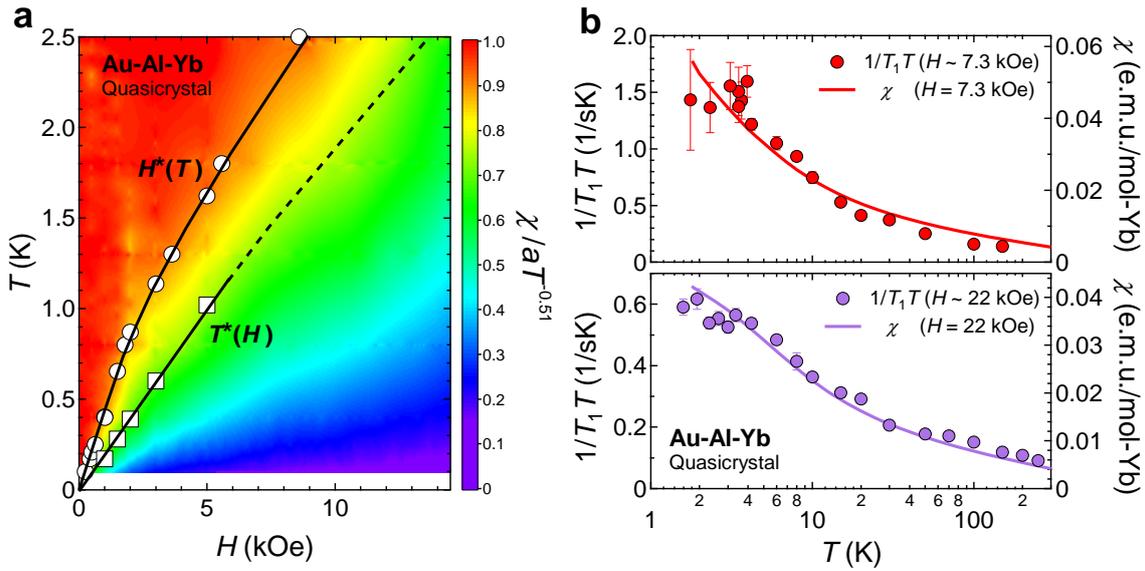

**Figure 3 | Magnetic properties of the quasicrystal. a**, Contour plot of the normalized uniform susceptibility $\chi(T,H)/\chi(T,0) = \chi(T,H)/aT^{-0.51}$ with $a = 0.081$ e.m.u.$K^{0.51}$mol$^{-1}$. The open circle and squares denote a crossover field $H^*$ defined by $\chi(T,H^*)/\chi(T,0) = 0.95$ and a characteristic temperature $T^*$, respectively. Solid lines are guides to the eye and the dashed line is an extrapolation along a contour line. **b**, Nuclear spin-lattice relaxation rate divided by temperature $1/T_1T$ of $^{27}$Al NMR (left scale) and the magnetic susceptibility $\chi(T)$ (right scale). The scaling relation $1/T_1T \propto \chi(T)$ is obviously observed. The abscissa is plotted in a logarithmic scale. The error bars of $1/T_1T$ indicate the standard errors from least-squares fits of the recovery data of nuclear magnetization.

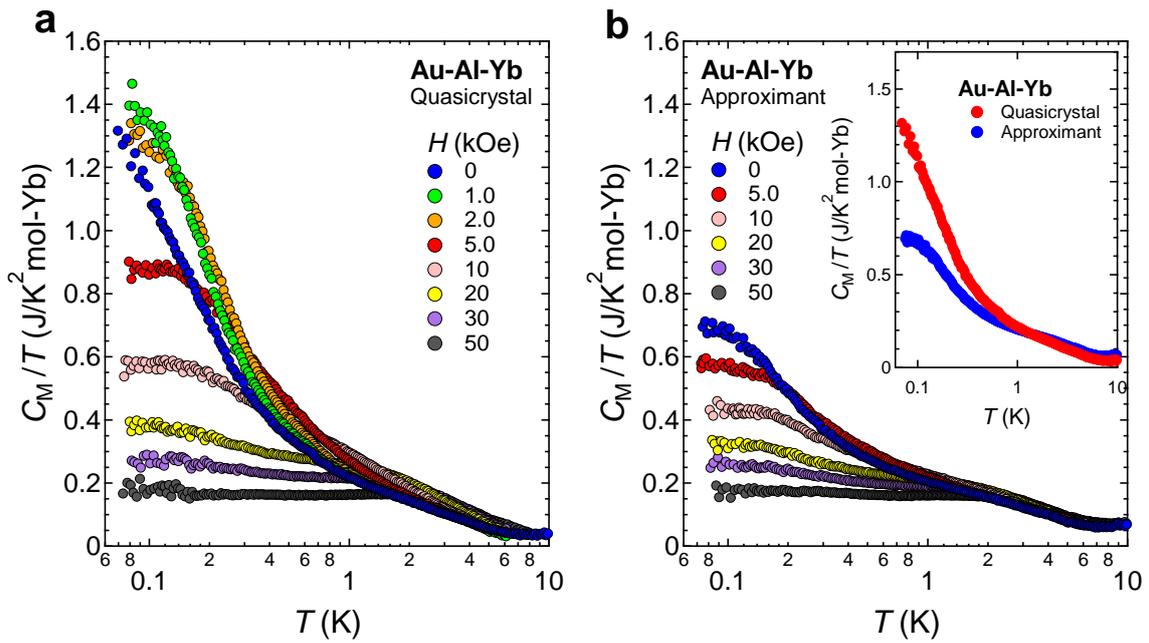

**Figure 4 | Temperature dependence of the magnetic specific heat $C_M/T$ of the quasicrystal and the approximant under magnetic field.** $C_M$ was obtained by subtracting the nuclear contribution, $C_N = \alpha(H)/T^2$, from the measured data, where $\alpha(H)$ was deduced from the plot of $CT^2$ vs $T^3$. **a**, Temperature dependence of $C_M/T$ of the quasicrystal. Note that $C_M/T$ diverges logarithmically at $H = 0$ whereas it tends to saturate under a magnetic field. **b**, $C_M/T$ of the approximant measured in the same condition as in Fig. 4a. Inset shows a comparison of $C_M/T$ between the quasicrystal and the approximant at zero field.

Supplemental Information for

**"Quantum critical state in a magnetic quasicrystal"**

Kazuhiko Deguchi, Shuya Matsukawa, Noriaki K. Sato,

Taisuke Hattori, Kenji Ishida, Hiroyuki Takakura & Tsutomu Ishimasa

**I.   Materials and methods**

Two kinds of alloy ingots were prepared using an arc furnace in an argon atmosphere; one has nominal composition of $Au_{49}Al_{34}Yb_{17}$, and the other $Au_{49}Al_{36}Yb_{15}$. The latter ingot was subsequently annealed at 700ºC for 24 h. Small pieces taken from each ingot were characterized by means of powder X-ray diffraction method, electron microscopy, and electron probe micro-analysis (EPMA) with a wave-length dispersive type spectrometer. The quasicrystal was prepared from the former composition ingot and found to include exclusively the icosahedral quasicrystal with the composition of $Au_{51}Al_{34}Yb_{15}$. The approximant was prepared from the latter composition ingot, and found to include exclusively the phase with the composition of $Au_{51}Al_{35}Yb_{14}$. Further details of the sample preparation are described in Ref. S1.

Four-terminal resistivity measurements were made using ac and dc methods down to 0.1 K. The magnetization was measured using a commercial SQUID magnetometer in the temperature range between 1.8 and 300 K. The ac magnetic susceptibility was measured using a driving field of a frequency of 100 Hz and a magnitude of 0.1 Oe down to $T = 0.1$ K. High pressure was generated using a piston cylinder clamped pressure cell with Daphne 7373 oil as a pressure transmitting medium. The applied pressure was estimated by measuring a superconducting transition of indium. The specific heat was measured by a relaxation method down to $T = 0.1$ K installed in a dilution refrigerator.

## II. High-temperature resistivity and magnetic susceptibility

Figure S1a shows the temperature dependence of the electrical resistivity $\rho(T)$ of the quasicrystal and the approximant. Both show metallic behaviour in a wide temperature range. The difference between the two systems emerges in the low temperature region: While the approximant indicates the Fermi liquid feature $\Delta\rho \propto T^2$ as in a conventional metal (inset of Fig. S1b), the quasicrystal exhibits the non-Fermi liquid behaviour $\Delta\rho \propto T$. Here, $\Delta\rho = \rho(T) - \rho(0)$ defines the temperature dependent part of the resistivity.

Figure S1b shows the temperature dependence of the uniform magnetic susceptibility. Above 100 K, both the quasicrystal and the approximant show a Curie–Weiss law, $\chi(T)=C/(T-\theta_W)$, where $C=N_A\mu^2_{\text{eff}}/3k_B$ is a Curie constant with the Avogadro number $N_A$ and the Boltzmann constant $k_B$, $\mu_{\text{eff}}$ is an effective moment, and $\theta_W$ is a Weiss temperature. These parameters are slightly different; $\mu_{\text{eff}} = 3.91\mu_B$ and $\theta_W = -153$ K for the quasicrystal, and $\mu_{\text{eff}} = 3.96\mu_B$ and $\theta_W = -117$ K for the approximant. Note that $\mu_{\text{eff}} = 4.54\mu_B$ and 0 for the free $Yb^{3+}$ and $Yb^{2+}$ ion, respectively.

## III. Quantum criticality and non-Fermi liquid

The low-temperature exponents of the physical properties obtained in the main text are summarized in Table S1. For comparison, the exponents of the Yb-based heavy fermion systems $YbAlB_4$ (Refs. S2 and S3) and $YbRh_2Si_2$ (Refs. S4-S6) are included together with those predicted by theories of the quantum valence criticality[S7] and the local quantum criticality[S8,S9]. We find the present criticality to resemble the exponents of YbAlB$_4$ and YbRh$_2$Si$_2$. The quantum criticality of these prototypical heavy fermions has been discussed based on the valence criticality model and the local criticality model. For the present system, our criticality coincides with the valence criticality. However, this model is unlikely to account for the observed robustness against pressure.

Quasicrystals are generally distinct from disordered systems: such inhomogeneous systems do not produce a clear, sharp diffraction pattern as shown in Fig. 1a. However, our quasicrystal is not ideal and probably includes imperfections such as the chemical disorder of Au and Al around Yb. Therefore, it may be possible to relate the present quantum criticality to non-Fermi liquid behaviour due to Kondo disorder and Griffiths phase in inhomogeneous disordered system[S10,S11].

According to these scenarios, however, it is expected that the electrical resistivity increases with lowering temperature and exhibits a negative magnetoresistance: these are inconsistent with our observation as confirmed in Fig. S2a. Furthermore, the absence of the power law dependence of the specific heat (right inset of Fig. S1b) and the robust exponent of the susceptibility against pressure, are both contrary to the disorder scenario predicting that $\chi \propto C/T \propto T^{\lambda-1}$ ($\lambda$ is a non-universal parameter depending on the hybridization). In a microscopic viewpoint, nuclear magnetization recoveries after saturation pulses (Fig. S2b), which were taken for the $1/T_1T$ measurements, were fit by the theoretical function with a single component of $T_1$ in the whole temperature range measured. This suggests that the electronic state in the quasicrystal Au-Al-Yb is rather homogeneous, which is contrasted with Kondo-disordered systems. As a consequence, our results are not explained by existing theories developed for crystalline materials, to our knowledge. We consider that the present results are helpful for developing a theory of quantum criticality and non-Fermi liquid in quasi-periodic systems.

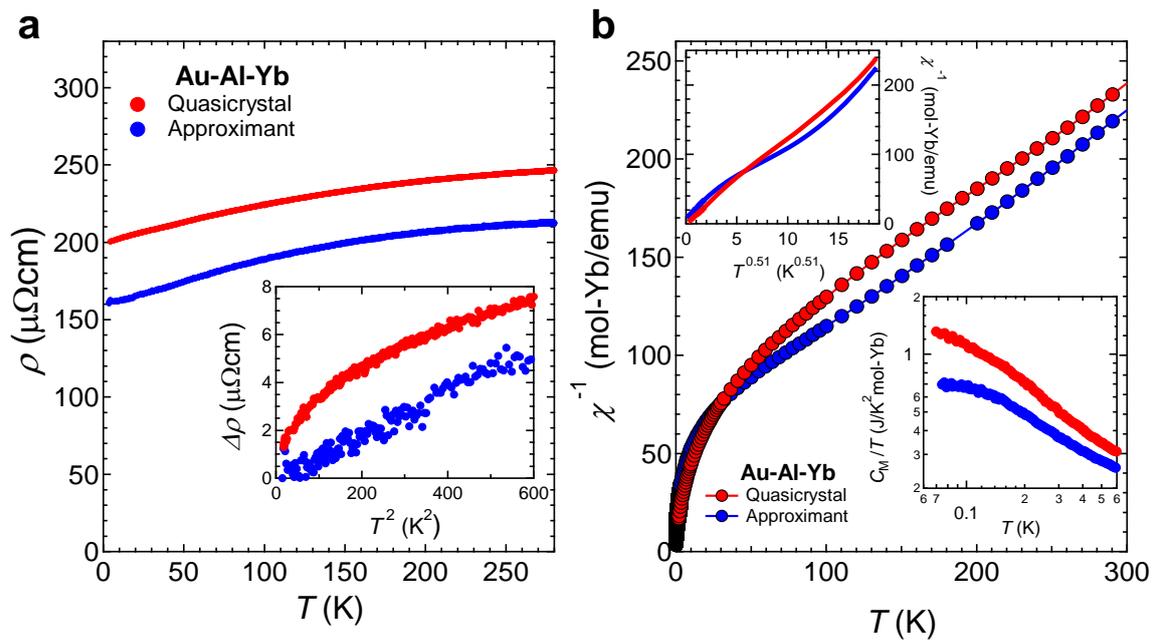

**Figure S1 | Electrical resistivity and magnetic susceptibility of the quasicrystal and the approximant. a**, Temperature dependence of the electrical resistivity $\rho$ at zero field. The residual resistivity is $\rho(0)$ = 199 and 161 $\mu\Omega$cm for the quasicrystal and the approximant, respectively. Inset shows the temperature dependent part $\Delta\rho = \rho(T) - \rho(0)$ as a function of $T^2$. **b**, Temperature dependence of the inverse of the uniform susceptibility $\chi^{-1}$. Left inset shows the inverse susceptibility $\chi^{-1}$ versus $T^{0.51}$ of the quasicrystal and the approximant below 300 K. Right inset shows the magnetic specific heat of the quasicrystal and the approximant on the double-log scale, respectively.

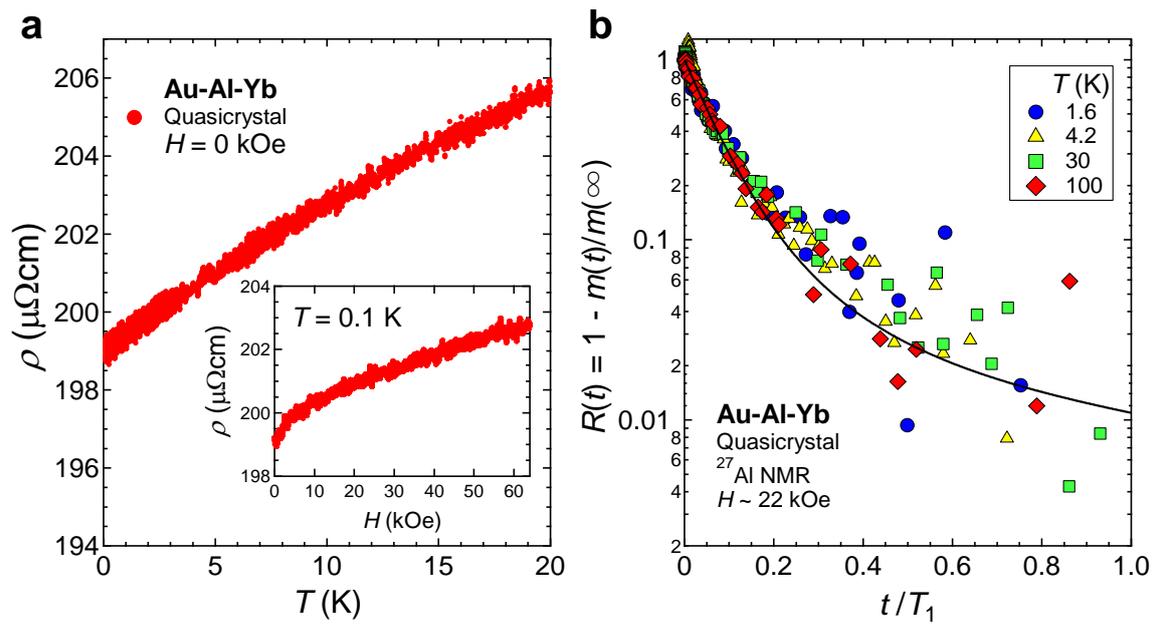

**Figure S2 | Electrical resistivity and the recovery curve of the nuclear magnetization of the quasicrystal. a**, Temperature dependence of the electrical resistivity $\rho$ at zero field down to 0.1 K. Inset shows magnetic field dependence of electrical resistivity $\rho$ at 0.1 K. **b**, The recovery of the nuclear magnetization after a saturation pulse of the quasicrystal. The data are well fit by a theoretical curve with a single component $T_1$ denoted by the solid line.

|  | Au-Al-Yb Quasicrystal | Au-Al-Yb Approximant | $\beta$-YbAlB$_4$ | YbRh$_2$Si$_2$ | Valence Criticality | Local Criticality | Kondo disorder Griffiths phase |
|---|---|---|---|---|---|---|---|
| $\chi^{-1}$ | $T^{0.51}$ | $T^{0.51}+$const. | $T^{0.5}$ | $T^{0.6}$ | $T^{0.5\sim0.7}$ | $T^{\alpha}+$const. ($\alpha<1$) | $T^{\lambda-1}$ ($0<\lambda\leq1$) |
| $1/T_1T$ | $\propto \chi$ | – | – | $T^{-0.5}$ | $\propto \chi$ | $T^{-1}$ | – |
| $C/T$ | $-\ln T$ | const. | $-\ln T$ | $-\ln T$ | $-\ln T$ | – | $\propto \chi$ |
| $\Delta\rho$ | $T$ | $T^2$ | $T^{1.5}\rightarrow T$ (low-$T$) | $T$ | $T$ | – | $-T$ ($\lambda$: non–universal parameter) |

**Table S1 | Exponents of some Yb-based materials together with those predicted by theories.** $\chi^{-1}$ is the inverse of the uniform magnetic susceptibility, $1/T_1T$ the nuclear spin-lattice relaxation rate divided by temperature, $C/T$ the specific heat divided by temperature, and $\Delta\rho = \rho(T) - \rho(0)$ the low-temperature resistivity. For comparison, the exponents of the periodic system YbRh$_2$Si$_2$ and $\beta$-YbAlB$_2$ are included. Our results for the Au-Al-Yb quasicrystal coincide with the prediction by the valence criticality theory.